# Operational approach to the entanglement.

Marian Kupczynski

*Département de l'Informatique, UQO, Case postale 1250 , succursale Hull, Gatineau. Québec, Canada J8X 3X 7*

**Abstract .** In early days of quantum theory  it was believed that the results of measurements performed on two distant physical systems should be uncorrelated thus their quantum state should be separable it means described  by a simple  tensor product of the individual local state vectors or a tensor product of individual local density operators.  It was shown many years ago by EPR that two systems which interacted in the past and separated afterwards have to be described in most cases by particular non-separable states which are called entangled.  It was noticed by Zanardi et al. that a Hilbert space of possible state vectors of compound physical system could be partitioned in different way by introducing various tensor product structures induced by the experimentally accessible observables (interactions and measurements). Therefore a separable state in one partition could become entangled in different partition. In this sense the entanglement was relative to a particular set of experimental capabilities.  Continuing this line of thought Torre  et al. claimed to prove that for any separable state there existed strongly correlated physical observables therefore  all quantum states were entangled. The same claims one may find in several recent papers.  In this paper we will discuss in operational way the differences existing between separable, non-separable and entangled states and we will show that the conclusions of Torre et al. were unjustified. A sufficient condition for entanglement is the violation of BI-CHSH and/or steering inequalities. Since there exist experiments outside the quantum physics violating these inequalities therefore in the operational approach one cannot say that the entanglement is an exclusive quantum phenomenon.

1. Introduction

The entanglement of quantum states is considered to be an important resource in Quantum Information therefore it should be well understood.  According to quantum theory (QT) the individual results of the measurements are obtained in irreducibly random way thus it seemed that the results of two random experiments measuring the physical observables A and B in distant locations should be independent. If it was true a quantum system studied should be necessarily described by a tensor product of some quantum states of its subsystems. Therefore  the results of local experiments should be uncorrelated  namely $E(A,B)= E(A)E(B)$  or in other words the covariance  $cov(A,B)$ should vanish .

In 1935 Einstein et al. (EPR) [1] demonstrated that the results of local measurements on two quantum systems which interacted in the past and separated afterwards were strongly correlated and had to be described by a particular non-separable state which was called entangled by Schrödinger [2].

 If two or several systems do interact forming a multi-partite system then various degrees of freedom of this system can be strongly correlated and the system is in so called generalised entangled state in contrast to the entangled state for non- interacting systems.   It was noticed by Zanardi et al.  [3,4] that a Hilbert space of possible state vectors of these physical systems could be partitioned in different way by introducing various tensor product  structures (TPS) induced by the experimentally accessible observables (interactions and measurements). Similar conclusions were reached by Barnum et al. [5]. In this sense the entanglement was relative to a particular set of experimental capabilities.

Continuing this line of thought Torre et al. [6] claimed to prove that any separable state was in fact entangled. They used the non-vanishing of quantum covariance function (QCF) as a criterion for the entanglement. Starting from separable quantum states and uncorrelated local observables *A* and *B* they showed that it was easy to find two functions *F(A,B)* and *G(A,B)* such that their QCF was non-zero and concluded that the entanglement was a universal property of any quantum state the point of view shared in several recent papers [7, 8].

First of all after defining in operational way non-separable and entangled states we will recall that the non-vanishing of covariance between some observables is not a sufficient condition for proving the entanglement. The entanglement can be only proven if having sufficient experimental capabilities one may demonstrate the violation of BI-CHSH inequalities and /or steering inequalities [9-13]. The classification of entangled states is not easy because there exist entangled mixed quantum states which do not violate BI-CHSH or steering inequalities [10, 16]. For more complicated multipartite systems one has to use so called entanglement witnesses [14, 15].

Next we will show by analyzing the examples from [6] that the functions *F(A,B)* and *G(A,B)* do not correspond to any new directly measurable observables and that their values can be only calculated using the observed values of uncorrelated observables *A* and *B*. Therefore we may say that the experimental capabilities [3-5] do not allow a new TPS in the Hilbert space induced by *F* and *G*. Since *F(A,B)* and *G(A,B)* are in general dependent random variables therefore the non-vanishing of their covariance is obvious and by no means can be considered as a proof that a separable quantum state is in fact entangled.

Finally in order to make even clearer that the non- vanishing of covariance has nothing to do with the entanglement we will give a simple example of two local random experiments with pairs of fair dices in distant locations giving the correlated outcomes.

2. **Operational definition of the entanglement.**

Let us consider an ensemble of identically prepared pairs of physical systems on which we can perform local coincidence measurements of some local physical observables *A* and *B*. Our ensemble can be two beams of "particles" sent by some source *S* or an ensemble of pairs of "quantum dots" obtained by resetting of the state of a particular pair of "quantum dots" before each repetition of the local measurements etc. The outcomes of local experiments have in general a statistical scatter thus can be interpreted as results of measurements of some random variables *A* and *B* for which we may define the expectation values and the covariance. If we use QT to describe our experiments we introduce a state vector $\psi \in H_1 \otimes H_2$ or a density operator $\rho$. In the discrete case local observables are represented by hermitian operators $\hat{A}_1 = \hat{A} \otimes I$ and $\hat{B}_1 = I \otimes \hat{B}$. Using this notation we define conditional expectation values: $E(A|\psi) = \langle \psi, \hat{A}_1 \psi \rangle$ or $E(A|\psi) = Tr\rho\hat{A}_1$ etc. We will use in the following the second form which is more general because it applies also to mixed quantum ensembles.

A conditional covariance of *A* and *B* is defined by:

$$\text{cov}(A,B|\rho) = E(AB|\rho) - E(A|\rho)E(B|\rho) \tag{1}$$

The conditional covariance of A and B coincides with quantum covariance function (QCF) used in [6]. Since the random variables A and B are independent thus the covariance function defined by Eq.1 has the obvious property:

$$\operatorname{cov}(kA+nB, mA+lB\,|\,\rho) = km\operatorname{var}(A\,|\,\rho) + nl\operatorname{var}(B\,|\,\rho) \tag{2}$$

Using the above introduced notation we may define separable, non- separable and entangled quantum states of two distant physical systems:

A state $\rho$ is separable if it can be written in the form $\rho = \rho_1 \otimes \rho_2$ where $\rho_1$ and $\rho_2$ are density operators acting in the Hilbert spaces $H_1$ and $H_2$ respectively. For separable states $\operatorname{cov}(A, B\,|\,\rho)$ vanish for all measurable pairs of local observables (A,B).

A state $\rho$ which is a convex sum of separable states: $\rho = \sum_{i=1}^{k} p_i \rho_i \otimes \tilde{\rho}_i$ with probabilities $0 < p_i < 1$ is nether neither separable nor entangled and we will call it non-separable. Since

$$E(AB\,|\,\rho) = \sum_{i=1}^{k} p_i E(A\,|\,\rho_i) E(B\,|\,\tilde{\rho}_i) \tag{3}$$

if the local expectation values are both different from zero then $\operatorname{cov}(A, B\,|\,\rho)$ does not vanish. Besides if $|E(A\,|\,\rho_i)| \leq 1$ and $|E(B\,|\,\tilde{\rho}_i)| \leq 1$ one can easily prove CHSH inequalities:

$$|E(AB\,|\,\rho) - E(AB'\,|\,\rho)| + |E(A'B\,|\,\rho) + E(A'B'\,|\,\rho)| \leq 2 \tag{4}$$

Therefore the correlations for a convex sum of separable states can be reproduced by so called local stochastic hidden variable (LHV) models [9].

A state $\rho$ is entangled if it cannot be written as a convex sum of separable states. It can be checked using quantum state tomography, entanglement witnesses or by showing that the expectation values for some pairs of measurable local observables violate Bell-CHSH or steering inequalities what in the operational approach provides the sufficient proof for the entanglement. Since a probabilistic LHV model is particularly suited to describe a convex sum of separable states (which we call non-separable) it is not strange that it does not provide a correct probabilistic model for the experiments with entangled states [17-25,28].

3. **Not all quantum states are entangled**

In their first example the authors [6] consider a quantum system described by a separable quantum state $\Psi$ depending on two space coordinates. The local position observables $X_1$ and $X_2$ are represented by $\hat{X}_1 = \hat{X} \otimes I$ and $\hat{X}_2 = I \otimes \hat{X}$ where $\hat{X}$ is a position operator. Of course $[\hat{X}_1, \hat{X}_2] = 0$ and $\operatorname{cov}(A, B\,|\,\psi) = 0$. Next they introduce two other variables $A = X_1 + X_2$ and $B = X_1 - X_2$. Using a similar formula to one given in Eq.2 they prove that $\operatorname{cov}(A, B\,|\,\psi)$ is equal to the difference of the variances of A and B which is in general different from zero and conclude that the separable state $\Psi$ is in fact entangled. Since it is impossible to measure the observables A and B directly therefore these observables do not introduce a new TPS in the Hilbert space of states and the separable quantum state does not become non-separable.

The same argument applies to their second example in which they consider two free spin $\frac{1}{2}$ particles in a separable state $\Psi$. They define new observables $S_z^2 = (S_z \otimes I + I \otimes S_z)^2$ and $S_x^2 = (S_x \otimes I + I \otimes S_x)^2$ having non vanishing covariance. Since the values of these observables can be only deduced from local measurements of the spin projections $S_z$ and $S_x$ thus the initial quantum stated does not become non-separable.

The authors mention also different TPS for quantum states of the hydrogen atom. It is true that the notion of entanglement was generalized in order to describe the coupling of different degrees of freedom of a single compound quantum system [4, 5] for example the hydrogen atom and is called a generalized entanglement. In this approach a quantum state of a system is called "entangled" if it is a non -convex sum of tensor products of some vectors representing different degrees of freedom of the compound system. For example for the hydrogen atom we find a natural factorization *CM-R* where *CM* denotes a center- of -mass and *R* relative motion degrees of freedom. Thus in *CM-R* splitting the quantum state is "separable".  If we wanted to use the degrees of freedom of the proton and the electron, so called *e-p* splitting, then formally the state vector of hydrogen atom becomes "entangled" in this new degrees of freedom. However the "center-of -mass+ relative degrees of freedom" structure appears as primarily operable form of the experimental reality of atoms" [7]. The e-p structure becomes useful when the atom ionizes. Let us note that talking about a "separable" quantum state in case of CM-R splitting of a state vector of an atom takes us far from the original idea of a separable state describing two non-interacting distant physical systems [1, 2] and may easily lead to a confusion. If we stick to the operational definitions of separable and entangled sates we stay on a safe ground.

4. **"Non-separable" ensemble of dices.**

Let us consider a following probabilistic random experiment. Carol can send pairs of dices to Alice and Bob. She has two types of fair dices D1 with 1 written on three faces and 0 on the remaining three faces and D2 with 1 written on 4 faces and 0 on the remaining two faces. She chooses to send a pair  (D1, D1) with the ``probability`` 0.25 and a pair (D2,D2) with the ``probability`` 0.75. Alice and Bob roll received dices and record their observations 0 or 1 and compare them. Using the language of mathematical statistics they measure the values of the corresponding random variables *A* and *B* on some mixed classical statistical ensemble.  It is easy to see that *E(A)=E(B)=E(A,B)=*0.625 thus the *cov(A,B)≠0* and if we  used a "quantum like" model [22] with  (2x2) density operators we would conclude that we have to use a non-separable "quantum like" state which of course is not entangled. Following [6] we could introduce new classical random variables *F(A,B)* and *G(A,B)* without gaining any new information about the "quantum like "state of the ensemble of dices sent by Carol.

5. **Conclusions**

In the operational approach the entanglement is an objective property of an ensemble of various expectation values *E(A,B)* found in coincidence  experiments performed on some identically prepared physical systems in different experimental settings. As we did show above not all quantum states are entangled.

To prove whether physical systems are prepared in an entangled quantum state one has to show that not all correlations between available local variables can be explained by using a convex sum of separable quantum states what can be proven if the BI-CHSH or steering inequalities are violated. Due to the ambiguities related to the finite statistics, efficiency of detectors, widths of the coincidence windows, post selection, noise etc. it is a difficult but not the impossible task to accomplish [11-13].

There exist many random experiments from outside the domain of quantum physics in which BI-CHSH are violated [22] therefore the entanglement as defined operationally is not exclusively a quantum phenomenon.

It is not strange since BI-CHSH may be interpreted as the necessary conditions for the existence of the joint probability distributions of the values of several dichotomous random variables which can be measured pair-wise but not simultaneously [17-25, 28].

One can even construct simple experiments in classical mechanics [28] in which some classical systems obey the momentum-energy conservation laws and the correlations between particular deduced binary observables violate BI.

It is also not well known but it is possible to simulate in a consistent and local way many experiments from quantum optics and neutron interferometry [26, 27] including those violating BI-CHSH .